\documentclass[a4paper,twocolumn,english]{revtex4}
\usepackage{graphicx}
\usepackage{amssymb}

\makeatletter

\usepackage{babel}
\makeatother
\begin{document}

\title{Casimir force between planes as a boundary finite size effect}

\author{Z. Bajnok$^{1}$, L. Palla$^{2}$, and G. Tak{\'a}cs$^{1}$}

\affiliation{$^{1}$\emph{\small Theoretical Physics Research Group, Hungarian
Academy of Sciences, 1117 Budapest, P{\'a}zm{\'a}ny P{\'e}ter s{\'e}t{\'a}ny 1/A, Hungary}
\\
$^{2}$\emph{\small Institute for Theoretical Physics, E{\"o}tv{\"o}s University,
1117 Budapest, P{\'a}zm{\'a}ny P{\'e}ter s{\'e}t{\'a}ny 1/A, Hungary}}

\begin{abstract}
The ground state energy of a boundary quantum field theory is derived
in planar geometry in D+1 dimensional spacetime. It provides a universal
expression for the Casimir energy which exhibits its dependence on
the boundary conditions via the reflection amplitudes of the low energy
particle excitations. We demonstrate the easy and straightforward
applicability of the general expression by analyzing the free scalar
field with Robin boundary condition and by rederiving the most important
results available in the literature for this geometry.
\end{abstract}
\maketitle

\section{Introduction}

The Casimir force between two neutral macroscopic bodies in vacuum
is considered as a manifestation of zero point fluctuations intrinsic
to any quantum field theory (QFT). It has received a lot of interest
recently both from the theoretical and from the experimental side
(cf. the recent reviews \cite{reviews} and the book by Milton \cite{book}). 

This force can be determined from the volume dependence of the ground
state energy (Casimir energy). It is usually derived either by a summation
of zero point fluctuations of the quantum field restricted to the
domain in question, or by analyzing the statistical fluctuations of
the interactions between the boundaries. Our new derivation is similar
to the first approach in the sense that it uses the quantum fields
living between the boundaries. However, we describe the quantum field
theory in question in terms of its particle excitations rather than
the field fluctuations. As a consequence we describe the Casimir energy
as 'virtual particle' exchange between the two boundaries and our
formula is expressed in terms of the analytically continued reflection
amplitudes of these particles. 

The interaction between the quantum field and the macroscopic bodies
is most often described by imposing appropriate boundary conditions
on the field. The aim of this paper is to explore and make explicit
the dependence of the Casimir force on the boundary conditions obeyed
by the field(s) in a geometry when the boundary conditions are given
on two parallel planes. We achieve this by regarding the force as
a physical manifestation of finite size effects in a \textsl{boundary}
QFT (BQFT).

In QFT, finite size effects in a periodic box were considered by L\"{u}scher
\cite{luscher}, who showed how the correction to particle masses
(infinite volume energies) can be expressed in terms of infinite volume
scattering data ($S$ matrix elements). Following this general idea,
we have recently determined the finite size correction to the ground
state energy in $2$ dimensional BQFT in terms of scattering data
on the half line (reflection amplitudes). This explicitly describes
how the ground state energy depends on the volume and it is obvious
from the results in \cite{bluscher} that a generalization to BQFT
in any spacetime dimension is straightforward and leads to a description
of the Casimir effect.

Boundary conditions in quantum field theory can be equally well described
by a boundary state in a crossed channel. This formalism was worked
out in detail for integrable 1+1 dimensional BQFTs \cite{GZ}, but
it can be extended to nonintegrable field theories in any number of
spacetime dimensions. We show that this formalism allows an alternative
method to derive the Casimir energy, and it has the advantage of using
only renormalized and phenomenologically meaningful field theoretic
quantities. A further advantage of using the boundary state formalism
is that it provides a systematic (large volume) expansion for the
Casimir energy in the interacting case.

Our main result is a large volume expression for the Casimir energy/force,
valid in any BQFT, which depends only on the reflection amplitudes
on the boundaries (although the boundary conditions considered allow
transmission as well). These reflection amplitudes can be determined
in a straightforward way in the half-infinite (one boundary) geometry,
and using them in the expression for the Casimir energy makes it unnecessary
to carry out the finite volume quantization separately in the various
applications.

The paper is organized as follows: in Section II we introduce BQFTs
and derive our main results for the volume dependence of their ground
state energy using the boundary state formulation. Section III contains
a description of several applications of the result in various BQFTs:
First we derive a formula for the Casimir energy of a free massive
boson subject to Robin type boundary condition. This expression is
tested against the well-known Dirichlet, Neumann and massless limits.
We further check our formula by recomputing the Casimir effect for
parallel dielectric slabs separated by a vacuum slot and for a massless
fermion subject to {}``bag boundary condition''. We make our concluding
remarks in Section IV. In the Appendix we confirm our result in a
simple case by presenting the naive derivation based on mode summation
tailor made for the framework of boundary QFTs.

\section{Derivation of the main formula }

Here we summarize the main aspects of BQFTs and present the novel
derivation of the Casimir energy. We follow the description of BQFTs
along the line of \cite{bLSZ} extending the result of \cite{GZ}
for higher dimensions. In the process we show that this formalism
makes it possible to establish a systematic (large volume) expansion
of the Casimir energy even for interacting fields.

\subsection{Boundary state formalism}

To simplify the presentation consider a quantum field theory of a
scalar field, $\Phi(t,x,\vec{y})$, in $D+1$ dimensions with Lagrangian\[
\mathcal{L}=\frac{1}{2}\left(\partial_{t}\Phi\right)^{2}-\frac{1}{2}\left(\partial_{x}\Phi\right)^{2}-\frac{1}{2}\left(\vec{\partial}\Phi\right)^{2}-V(\Phi)\]
where $\vec{y}$ is a $D-1$ dimensional position vector, $\vec{\partial}_{i}=\frac{\partial}{\partial y_{i}}$.
We also suppose that the spectrum consists of one particle type with
mass $m$. The Hilbert space of the model can be spanned by asymptotic
multi-particle states with momentum parameterized by $(k_{i},\vec{k}_{i})$
(where again $\vec{k}_{i}$ denotes a $D-1$ dimensional momentum
vector):\[
\mathcal{H}=\left\{ \vert k_{1},\vec{k}_{1};k_{2},\vec{k}_{2};\dots;k_{n},\vec{k}_{n}\rangle\right\} \]
The energy of a one-particle state is\[
\omega(k,\vec{k})=\sqrt{m^{2}+k^{2}+\vec{k}^{2}}\]
In the asymptotic scattering configurations (large negative or positive
time) the particles are distant from each other, and the spectrum
can be described by free \emph{in} or \emph{out} particles so the
energy of the multi-particle state is the sum of the individual energies.
These two Hilbert spaces are connected by the scattering matrix\begin{eqnarray*}
S(\{ k_{i}^{'},\vec{k}_{i}^{'}\},\{ k_{i},\vec{k}_{i}\}) & =\\
 &  & \hspace{-3cm}{}^{out}\langle k_{1}^{'},\vec{k}_{1}^{'};k_{2}^{'},\vec{k}_{2}^{'};...;k_{n}^{'},\vec{k}_{n}^{'}\vert k_{1},\vec{k}_{1};k_{2},\vec{k}_{2};...;k_{m},\vec{k}_{m}\rangle^{in}\end{eqnarray*}
 which can be expressed in terms of the correlators\begin{eqnarray*}
G(t_{1},x_{1},\vec{y}_{1};\dots;t_{N},x_{N},\vec{y}_{N}) & =\\
 &  & \hspace{-2cm}\left\langle 0\right|\Phi(t_{1},x_{1},\vec{y}_{1})\dots\Phi(t_{N},x_{N},\vec{y}_{N})\left|0\right\rangle \end{eqnarray*}
via the LSZ reduction formula. The time evolution of the state is
generated by the Hamiltonian \[
H=\int_{-\infty}^{\infty}dx\int d\vec{y}\left(\frac{1}{2}\Pi^{2}+\frac{1}{2}\bigl(\partial_{x}\Phi\bigr)^{2}+\frac{1}{2}\bigl(\vec{\partial}\Phi\bigr)^{2}+V(\Phi)\right)\]
where $\Pi=\partial_{t}\Phi$ is the conjugate momentum. 

Let us suppose now that the theory is restricted to the half space
$x<0$ and the boundary condition is given by specifying some boundary
potential at $x=0$\[
V_{B}(\Phi(0,\vec{y},t))\]
 The essential observation is that we can have two different Hamiltonian
descriptions of this system. We can take $t$ as the time variable
and then the boundary is situated in space (i.e. it has a spacelike
normal vector). In this case the Hilbert space consist of multi-particle
states\[
\mathcal{H}_{B}=\left\{ \vert k_{\perp},\vec{k}_{\parallel};k'_{\perp},\vec{k'}_{\parallel};\dots\rangle_{B}\right\} \]
where we indicate explicitly that the full momentum vector is composed
of two components, one perpendicular and the other parallel to the
boundary. The states are normalized in the following way: \begin{eqnarray*}
 &  & \,_{B}^{in}\langle k'_{\perp},\vec{k}'_{\parallel}\vert k_{\perp},\vec{k}_{\parallel}\rangle_{B}^{in}=\\
 &  & \left(2\pi\right)^{D}\omega(k_{\perp},\vec{k}_{\parallel})\delta\left(k_{\perp}-k'_{\perp}\right)\delta^{(D-1)}\left(\vec{k}_{\parallel}-\vec{k}'_{\parallel}\right)\end{eqnarray*}
The subscript $B$ indicates these states satisfy the boundary condition.
In the asymptotic past all the particles move towards the boundary:
$k_{\perp}>0$. For $t\to-\infty$ they are separated far from each
other and from the boundary forming the \emph{in} state, which is
a free state. For $t\to\infty$ all the scatterings and reflections
are terminated they move backwards from the boundary are far from
each other and from the boundary forming the free \emph{out} state.
These two Hilbert spaces are connected by the reflection matrix\[
R_{\alpha\beta}={}_{B}^{out}\langle\alpha\vert\beta\rangle_{B}^{in}\]
The simplest case describes the one particle elastic reflection on
the boundary and can be written as\begin{eqnarray*}
 &  & R(\theta,m_{\mathrm{eff}}(\vec{k}_{||}))\left(2\pi\right)^{D}\delta\left(\theta-\theta'\right)\delta^{(D-1)}\left(\vec{k}_{\parallel}-\vec{k}'_{\parallel}\right)=\\
 &  & {}_{B}^{out}\langle k_{\perp}^{'},\vec{k}_{||}^{'}\vert k_{\perp},\vec{k}_{||}\rangle_{B}^{in}\end{eqnarray*}
where we exploited the unbroken Poincar{\'e} symmetry (in the coordinates
$t,\:\vec{y}$) and parameterized the perpendicular momentum as\[
k_{\perp}=m_{\mathrm{eff}}(\vec{k}_{||})\sinh\theta\quad;\qquad m_{\mathrm{eff}}(\vec{k}_{||})=\sqrt{m^{2}+\vec{k}_{||}^{2}}\]
and then\[
\delta\left(\theta-\theta'\right)=m_{\mathrm{eff}}(\vec{k}_{||})\cosh\theta\,\delta\left(k_{\perp}-k'_{\perp}\right)\]
 The reflection factors can be expressed in terms of the correlators\begin{eqnarray*}
G(t_{1},x_{1},\vec{y}_{1};\dots;t_{N},x_{N},\vec{y}_{N}) & =\\
 &  & \hspace{-2cm}_{B}\left\langle 0\right|\Phi(t_{1},x_{1},\vec{y}_{1})\dots\Phi(t_{N},x_{N},\vec{y}_{N})\left|0\right\rangle _{B}\end{eqnarray*}
via the boundary reduction formula \cite{bLSZ}. The time evolution
of the state is generated by the Hamiltonian

\begin{eqnarray*}
H_{B} & = & \int_{-\infty}^{0}dx\int d\vec{y}\biggl[\frac{1}{2}\Pi^{2}+\frac{1}{2}\bigl(\partial_{x}\Phi\bigr)^{2}+\frac{1}{2}\bigl(\vec{\partial}\Phi\bigr)^{2}\\
 &  & \hspace{3cm}+V(\Phi)+\delta(x)V_{B}(\Phi)\biggr]\end{eqnarray*}
In the second description of the boundary theory we can take $\tau=ix$
as time and $\xi=-it$ as space variable in the Hamiltonian formalism
\footnote{In the Euclidean formalism this corresponds to swapping Euclidean
time with the space variable $x$, which is manifestly a symmetry
of the theory in the bulk.%
}. The Hilbert space now is that of the bulk theory (without any boundary),
and time evolution is given by the bulk Hamiltonian (this is the extension
of the crossing symmetry of the bulk theory, and is analogous to the
open-closed duality in string theory\emph{)}. In this case the boundary
condition appears in time and serves as an initial state in calculating
correlators:\begin{eqnarray*}
G(\tau_{1},\xi_{1},\vec{y}_{1};\dots;\tau_{N},\xi_{N},\vec{y}_{N}) & =\\
 &  & \hspace{-2cm}\left\langle 0\right|\Phi(\tau_{1},\xi_{1},\vec{y}_{1})\dots\Phi(\tau_{N},\xi_{N},\vec{y}_{N})\left|B\right\rangle \end{eqnarray*}
where now $\left|B\right\rangle $ is a state in the bulk Hilbert
space $\mathcal{H}$. The boundary state $\left|B\right\rangle $
is in fact defined by the equality of correlation functions calculated
in the two pictures. Using asymptotic completeness it can be expanded
in the basis of the asymptotic \emph{in} states of the bulk theory
\begin{eqnarray}
\left|B\right\rangle  & = & \left\{ 1+K_{1}A_{in}^{\dagger}(0,0)+\int_{0}^{\infty}\frac{d\theta}{2\pi}\int\frac{d^{D-1}\vec{k}_{||}}{\left(2\pi\right)^{D-1}}\right.\label{eq:B_expansion}\\
 &  & \left.\hspace{-.5cm}K_{2}\left(\theta,m_{\mathrm{eff}}(\vec{k}_{||})\right)A_{in}^{\dagger}(-\theta,-\vec{k}_{||})A_{in}^{\dagger}(\theta,\vec{k}_{||})+\dots\right\} \left|0\right\rangle \nonumber \end{eqnarray}
where $A^{\dagger}(\theta,\vec{k}_{||})$ is the \emph{in} asymptotic
creation operator and the ellipses denote the terms with higher particle
number. Due to translational invariance in the spatial direction all
the contributing states must have zero total momentum. Ghoshal and
Zamolodchikov have shown \cite{GZ} that the coefficient $K_{2}$
is related to the one-particle elastic reflection factor $R$ on the
boundary by \[
K_{2}\left(\theta,m_{\mathrm{eff}}(\vec{k}_{||})\right)=R\left(\frac{i\pi}{2}-\theta,m_{\mathrm{eff}}(\vec{k}_{||})\right)\]
(here we simply treat $\vec{k}_{||}$ as a label for infinitely many
species of two-dimensional particles of mass $m_{\mathrm{eff}}(\vec{k}_{||})$,
using the fact that $\vec{k}_{||}$ is conserved in the reflection
off the boundary). The one-particle coefficient $K_{1}$ is only nonzero
when the vacuum expectation value of the interpolating field $\Phi$
of the particle does not vanish. In this work we suppose that $\,_{B}\left\langle 0\right|\Phi\left(t,x,\vec{y}\right)\left|0\right\rangle _{B}=0$
(which is the case for the usual applications). We plan to present
the general case in a forthcoming publication, together with more
detailed derivation of the form of the coefficient $K_{1}$ and $K_{2}$
using the boundary reduction formula obtained in \cite{bLSZ}, and
a generalization to more complex particle spectra with several different
masses. Here we restrict ourselves to a vanishing $K_{1}$ to keep
the discussion simple. We remark that when the theory in the bulk
is free and the reflection is elastic, the boundary state can be written
in a closed form %
\footnote{In 1+1 dimensions this can be extended to any integrable QFT with
integrable boundary condition \cite{GZ}. %
}\begin{eqnarray}
\left|B\right\rangle  & = & \exp\biggl(\int_{0}^{\infty}\frac{d\theta}{2\pi}\int\frac{d^{D-1}\vec{k}_{||}}{\left(2\pi\right)^{D-1}}\label{eq:B_exp}\\
 &  & K_{2}\bigl(\theta,m_{\mathrm{eff}}(\vec{k}_{||})\bigr)A_{in}^{\dagger}(-\theta,-\vec{k}_{||})A_{in}^{\dagger}(\theta,\vec{k}_{||})\biggr)\left|0\right\rangle \nonumber \end{eqnarray}
We note also that if we have more then one particle species, with
the same mass $m$ created by $A_{in}^{\dagger}(\theta,\vec{k}_{||})_{j}$
(i.e. a multiplet), the formula for the boundary state changes as 

\begin{eqnarray*}
\left|B\right\rangle  & = & \biggl\{1+\int_{0}^{\infty}\frac{d\theta}{2\pi}\int\frac{d^{D-1}\vec{k}_{||}}{\left(2\pi\right)^{D-1}}\\
 &  & \hspace{-1cm}K_{2}^{ij}\bigl(\theta,m_{\mathrm{eff}}(\vec{k}_{||})\bigr)A_{in}^{\dagger}(-\theta,-\vec{k}_{||})_{i}A_{in}^{\dagger}(\theta,\vec{k}_{||})_{j}+\dots\biggr\}\left|0\right\rangle \\
\end{eqnarray*}
Boundary conditions considered in the context of the Casimir effect
generally allow transmission as well, and such boundaries are called
'defects'. A suitable generalization of the above formalism can be
obtained by a folding trick, which maps the defect into a boundary
system \cite{defect}. Suppose now that a defect is located at $x_{0}$.
In the crossed channel picture it can be represented by a defect operator
which acts from the bulk Hilbert space of the $x<x_{0}$ system into
that of the $x>x_{0}$ system. Lets denote the operator creating the
particle for the $x<x_{0}$ domain as $A_{1}^{\dagger}$ while for
the $x>x_{0}$ domain as $A_{2}^{\dagger}$. There are now four one-particle
reflection amplitudes: $R^{\pm}$ are the ones preserving the species
number $1,2$, while $T^{\pm}$ are the ones changing $1$ into $2$
and $2$ into $1$, respectively. Upon the folding correspondence,
$R^{\pm}$ correspond to reflection processes on the two sides of
the defect, while $T^{\pm}$ describe the transmission amplitudes
from one side to the other. Using the folding map to the boundary
system we obtain the defect operator \cite{defect} as \begin{eqnarray}
D & = & 1+\int_{\-\infty}^{\infty}\frac{d\theta}{4\pi}\int\frac{d^{D-1}\vec{k}_{||}}{\left(2\pi\right)^{D-1}}\biggl(\label{eq:D_expansion}\\
 &  & R^{+}\Bigl(\frac{i\pi}{2}-\theta,m_{\mathrm{eff}}(\vec{k}_{||})\Bigr)A_{1}^{\dagger}(-\theta,-\vec{k}_{||})A_{1}^{\dagger}(\theta,\vec{k}_{||})+\nonumber \\
 &  & \hspace{0cm}T^{+}\Bigl(\frac{i\pi}{2}-\theta,m_{\mathrm{eff}}(\vec{k}_{||})\Bigr)A_{1}^{\dagger}(-\theta,-\vec{k}_{||})A_{2}(-\theta,-\vec{k}_{||})+\nonumber \\
 &  & T^{-}\Bigl(\frac{i\pi}{2}-\theta,m_{\mathrm{eff}}(\vec{k}_{||})\Bigr)A_{1}(\theta,\vec{k}_{||})A_{2}^{\dagger}(\theta,\vec{k}_{||})+\nonumber \\
 &  & \hspace{0cm}R^{-}\Bigl(\frac{i\pi}{2}-\theta,m_{\mathrm{eff}}(\vec{k}_{||})\Bigr)A_{2}(\theta,\vec{k}_{||})A_{2}(-\theta,-\vec{k}_{||})\biggr)+\dots\nonumber \end{eqnarray}
which (with the same conditions as for the boundary state, but now
elasticity is required for the combined one-particle reflection/transmission
amplitude \cite{mussardo} %
\footnote{We remark that integrable defects (which are exactly the ones for
which the defect operator can be exponentiated) with nontrivial reflection
and transmission at the same time are only possible when the bulk
scattering is trivial \cite{mussardo}. This only means a restriction
in $1+1$ dimensions, where integrable theories with nontrivial bulk
scattering matrices are possible. In higher space-time dimension the
conditions for the exponential form of the defect operator include
the triviality of the bulk $S$ matrix anyway.%
}) can be exponentiated to the form\begin{eqnarray}
D & = & \exp\biggl\{\int_{\-\infty}^{\infty}\frac{d\theta}{4\pi}\int\frac{d^{D-1}\vec{k}_{||}}{\left(2\pi\right)^{D-1}}\label{eq:D_exp}\\
 &  & \left(R^{+}\Bigl(\frac{i\pi}{2}-\theta,m_{\mathrm{eff}}(\vec{k}_{||})\Bigr)A_{1}^{\dagger}(-\theta,-\vec{k}_{||})A_{1}^{\dagger}(\theta,\vec{k}_{||})+\right.\nonumber \\
 &  & \hspace{0cm}T^{+}\Bigl(\frac{i\pi}{2}-\theta,m_{\mathrm{eff}}(\vec{k}_{||})\Bigr)A_{1}^{\dagger}(-\theta,-\vec{k}_{||})A_{2}(-\theta,-\vec{k}_{||})+\nonumber \\
 &  & T^{-}\Bigl(\frac{i\pi}{2}-\theta,m_{\mathrm{eff}}(\vec{k}_{||})\Bigr)A_{1}(\theta,\vec{k}_{||})A_{2}^{\dagger}(\theta,\vec{k}_{||})\nonumber \\
 &  & \left.\hspace{0cm}+R^{-}\Bigl(\frac{i\pi}{2}-\theta,m_{\mathrm{eff}}(\vec{k}_{||})\Bigr)A_{2}(\theta,\vec{k}_{||})A_{2}(-\theta,-\vec{k}_{||})\right)\biggr\}\nonumber \end{eqnarray}

\subsection{Derivation of Casimir energy}

Let us now turn to the derivation of Casimir energy of a $D+1$ dimensional
scalar field $\Phi(t,x,\vec{y})$ in a domain of width $L$ in $x$.
To facilitate later applications, it is useful to consider the case
when the field $\Phi(t,x,\vec{y})$ is allowed to penetrate through
the ends of the domain in such a way, that in the two domains of width
$R$ adjoining $L$, the dispersion relation of the asymptotic particles
may be different from the vacuum one in $L$. The totally reflecting
boundaries at $x=\pm(R+L/2)$ are treated as auxiliary boundary conditions
which are necessary in order to have a discrete spectrum; at the end
of calculations we shall take the limit $R\rightarrow\infty$ and
check that the results are independent of the choice of the auxiliary
boundary conditions. So let us take the following situation in the
coordinate $x$

\begin{center}\includegraphics[%
  width=8cm,
  keepaspectratio]{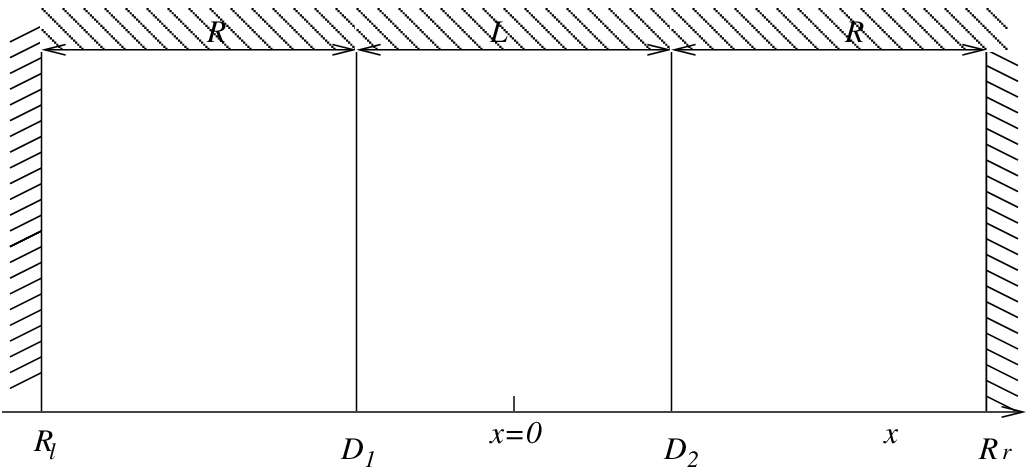}\end{center}

where total reflection occurs at $x_{l}=-R-L/2$ and $x_{r}=R+L/2$
with reflection factors $R_{l}(\theta,m_{\mathrm{eff}}(\vec{k}_{||}))$
and $R_{r}(\theta,m_{\mathrm{eff}}(\vec{k}_{||}))$ and there are
two {}``defects'' (i.e. boundaries allowing transmission) located
at $x_{1}=-L/2$ and $x_{2}=L/2$ with one-particle defect matrices
\[
D_{i}(\theta,m_{\mathrm{eff}}(\vec{k}_{||}))=\left(\begin{array}{cc}
R_{i}^{+}(\theta,m_{\mathrm{eff}}(\vec{k}_{||})) & T_{i}^{-}(\theta,m_{\mathrm{eff}}(\vec{k}_{||}))\\
T_{i}^{+}(\theta,m_{\mathrm{eff}}(\vec{k}_{||})) & R_{i}^{-}(\theta,m_{\mathrm{eff}}(\vec{k}_{||}))\end{array}\right)\]
 where $i=1,2$. The Hamiltonian of the system is denoted by $H_{B}$
and its Hilbert space by $\mathcal{H}_{B}.$ It is the ground state
energy of the system which is of interest for calculating the Casimir
effect, i.e. the lowest eigenvalue of $H_{B}$, which can be evaluated
via the partition function. Compactifying all infinite (temporal and
spatial) dimensions on circles with perimeter $T$ we can calculate
the partition function of our system in two different ways:\begin{eqnarray*}
Z_{R}(L,T) & = & \mathrm{Tr}_{\mathcal{H}_{B}}\mathrm{e}^{-TH_{B}}\\
 & = & \left\langle B_{l}\right|\mathrm{e}^{-RH_{x}^{(1)}}D_{1}\mathrm{e}^{-LH_{x}^{(2)}}D_{2}\mathrm{e}^{-RH_{x}^{(3)}}\left|B_{r}\right\rangle \end{eqnarray*}
 where $H_{x}^{(i)}$ is the periodic Hamiltonian in the $x$ channel
in the three domains indexed by $i=1,2,3$. Inserting complete sets
of bulk states, taking the limit $R\rightarrow\infty$ and normalizing
the ground state energy in infinite space to $E_{0}=0$ we obtain\begin{eqnarray*}
Z_{\infty}(L,T) & = & \sum_{n}\left\langle B_{l}\right|0\rangle\left\langle 0\right|D_{1}\left|n\right\rangle \left\langle n\right|D_{2}\left|0\right\rangle \langle0\left|B_{r}\right\rangle \mathrm{e}^{-LE_{n}}\\
 & = & \sum_{n}\left\langle 0\right|D_{1}\left|n\right\rangle \left\langle n\right|D_{2}\left|0\right\rangle \mathrm{e}^{-LE_{n}}\end{eqnarray*}
where the dependence on the auxiliary boundary conditions $B_{l,r}$
drops out (since in the $R\rightarrow\infty$ limit only the vacuum
state contributes from the expansions of $\left|B_{l,r}\right\rangle $).
The first few terms can be written explicitly as\begin{eqnarray*}
1+\sum_{\theta,\vec{k}_{||}}\sum_{\theta',\vec{q}_{||}}\left\langle 0\right|D_{1}\vert\theta,\vec{k}_{||};\theta',\vec{q}_{||}\rangle\langle\theta,\vec{k}_{||};\theta',\vec{q}_{||}\vert D_{2}\left|0\right\rangle \times\\
e^{-L\left(m_{\mathrm{eff}}(\vec{k}_{||})\cosh\theta+m_{\mathrm{eff}}(\vec{q}_{||})\cosh\theta'\right)}+O(e^{-3mL})\end{eqnarray*}
 The term $1$ is the contribution from the vacuum ($\left|n\right\rangle =\left|0\right\rangle $),
the next term comes from two-particle terms in (\ref{eq:D_expansion})
and the higher-order corrections come from the higher multi-particle
terms. This is a sort of cluster expansion similar to the one used
in \cite{bluscher}, valid for large values of the volume $L$. It
is obvious from these expressions that the leading (two-particle)
contribution depends only on $R_{1}^{-}$ and $R_{2}^{+}$. The ground
state (Casimir) energy (per unit transverse area) can be extracted
from the partition function as \[
E(L)=-\lim_{T\rightarrow\infty}\frac{1}{T^{D}}\log Z_{\infty}(L,T)\]
The result is \begin{eqnarray}
E(L) & = & -\int_{-\infty}^{\infty}\frac{d\theta}{4\pi}\,\cosh\theta\int\frac{d^{D-1}\vec{k}_{||}}{(2\pi)^{D-1}}m_{\mathrm{eff}}(\vec{k}_{||})\nonumber \\
 &  & \hspace{-1.3cm}R_{1}^{-}\Bigl(\frac{i\pi}{2}+\theta,m_{\mathrm{eff}}(\vec{k}_{||})\Bigr)R_{2}^{+}\Bigl(\frac{i\pi}{2}-\theta,m_{\mathrm{eff}}(\vec{k}_{||})\Bigr)\times\nonumber \\
 &  & \hspace{1cm}\mathrm{e}^{-2m_{\mathrm{eff}}(\vec{k}_{||})\cosh\theta L}+O(e^{-3mL})\label{eq:leading_order}\end{eqnarray}
The correction terms correspond to higher particle terms in the expansion
(\ref{eq:D_expansion}) of the defect operator $D$ and include the
amplitudes of reflection/transmission processes involving more than
one particle in at least one of the asymptotic states. These can be
computed e.g. using a BQFT formulation as the one in \cite{bLSZ},
but it is obvious that they are suppressed by a factor $\mathcal{\mathrm{e}}^{-mL}$
with respect to the leading order term due to the presence of at least
one additional particle in the corresponding term of the expansion
of the defect operator $D$. 

The formula (\ref{eq:leading_order}), describing Casimir effect to
leading order at long distances, is the main result of this paper.
Note that it is applicable in the presence of nontrivial bulk and
boundary interactions: their effects at leading order are contained
in the reflection factors $R^{\pm}$, so as long as there is some
theoretical or experimental input from which these can be determined
the leading order contribution can be evaluated. In integrable (2D)
boundary theories $R^{\pm}$ are obtained as solutions of the boundary
Yang Baxter equation, and the bulk interaction manifests itself thorough
the bulk S matrix appearing in this equation. In case of nonintegrable
interacting bulk theories e.g. a perturbative expansion can be given
to determine $R^{\pm}$, (for more details see \cite{bLSZ}).

Another important point is that this approach formulates the Casimir
effect from an infrared viewpoint. Standard derivations of the Casimir
effect (such as the one presented in the appendix) solve the microscopic
field theory. This necessitates tackling diverse issues such as renormalization,
and also the possibility that the infrared (long distance behaviour)
may be quite different from the microscopic description of the theory
(as is the case for example in QCD). Formula (\ref{eq:leading_order})
is expressed in terms of the asymptotic particles, the long distance
degrees of freedom, and provides a long distance expansion for Casimir
energy. It describes a finite size effect in a boundary quantum field
theory close in spirit to our previous investigation of the boundary
L{\"u}scher formula \cite{bluscher}, and it is not difficult to see that
the latter is just a special case of (\ref{eq:leading_order}). As
a consequence our formula has already been tested in interacting (integrable)
2D quantum field theories in \cite{bluscher}. (This is the only case
to our knowledge where exact reflection factors of interacting BQFTs
have been computed). Furthermore, the result (\ref{eq:leading_order})
includes the contribution of states localized to the defects (called
'surface plasmons' in the literature), since they are taken into account
as poles at imaginary rapidity of the reflection factors. This fact
is also demonstrated explicitely using the zero mode summation method
in Appendix A. 

A very appealing property of the boundary state approach is its universality
and we shall see in the next section that it indeed reproduces all
the results previously known for the planar situation. But in those
cases we can go even further, because in most calculations of the
Casimir effect the bulk is free and the boundary scattering is elastic,
therefore one can use the exponentiated boundary state (\ref{eq:B_exp})
together with the analogous defect operator (\ref{eq:D_exp}) which
makes it possible to sum up all the multi-particle terms of the cluster
expansion. The calculation to be done is essentially identical to
the derivation of the boundary Thermodynamic Bethe Ansatz equation
with trivial bulk $S$ matrix and reflection factors $R_{1}^{-}$
and $R_{2}^{+}$, which was already performed in \cite{lmss}. The
only difference is that \cite{lmss} supposes fermionic statistics
and that in our case the reflection factors $R_{1}^{-}$ and $R_{2}^{+}$
are not unitary due to the existence of transmission, but these do
not change the overall reasoning of the derivation. The result is\begin{eqnarray}
E(L) & = & \pm\int_{-\infty}^{\infty}\frac{d\theta}{4\pi}\,\cosh\theta\int\frac{d^{D-1}\vec{k}_{||}}{(2\pi)^{D-1}}m_{\mathrm{eff}}(\vec{k}_{||})\nonumber \\
 &  & \hspace{-1.5cm}\log\biggl(1\mp R_{1}^{-}\Bigl(\frac{i\pi}{2}+\theta,m_{\mathrm{eff}}(\vec{k}_{||})\Bigr)R_{2}^{+}\Bigl(\frac{i\pi}{2}-\theta,m_{\mathrm{eff}}(\vec{k}_{||})\Bigr)\nonumber \\
 &  & \hspace{3cm}\mathrm{e}^{-2m_{\mathrm{eff}}(\vec{k}_{||})\cosh\theta L}\biggr)\label{eq:alt3}\end{eqnarray}
(the upper/lower signs correspond to the bosonic/fermionic case, respectively).
Although formula (\ref{eq:leading_order}) is more general and adequate
enough for the large distance regime, the theoretical situations considered
in the literature can be directly compared to (\ref{eq:alt3}). It
is obvious that for $L\gg m^{-1}$ (\ref{eq:alt3}) reduces to (\ref{eq:leading_order}).
The Casimir force is dominated by modes with momentum of the order
$1/L$; for large enough separation the bulk interaction can be dropped
due to its short-ranged nature, and at the same time the characteristic
energy is lower than the threshold for inelastic boundary scattering.

\section{Applications}

To show the universal nature of these results we consider some applications.
We demonstrate that taking the reflection factors from the literature
and substituting into (\ref{eq:alt3}) immediately gives the Casimir
energy without going through a detailed analysis of the microscopic
quantum fluctuations.

As a first application we determine the Casimir energy for a \textsl{massive}
free scalar field, $V(\Phi)=\frac{m^{2}}{2}\Phi^{2}$, subject to
Robin boundary conditions, $V_{B}(\Phi)=\frac{c}{2}\Phi^{2}$, on
two parallel hyperplanes. If the two planes are located at $x=0$
and $x=L$ the boundary conditions are given as \[
\partial_{x}\Phi-c_{1}\Phi|_{x=0}=0;\quad\partial_{x}\Phi+c_{2}\Phi|_{x=L}=0;\quad c_{1},c_{2}\geq0,\]
 and the reflection amplitudes on these planes can be written as \[
R_{j}(\theta,m_{\mathrm{eff}}(\vec{k}_{||}))=\frac{m_{\mathrm{eff}}(\vec{k}_{||})\sinh\theta-ic_{j}}{m_{\mathrm{eff}}(\vec{k}_{||})\sinh\theta+ic_{j}};\quad j=1,2.\]
 Note that they have the same form as in the two dimensional case,
but now they depend also on $\vec{k}_{||}$ via the rapidity parameterization
$k_{\perp}=m_{\mathrm{eff}}(\vec{k}_{||})\sinh\theta$. When $\theta$
is continued to $\theta\rightarrow\theta+i\frac{\pi}{2}$, we have
\[
R_{j}\bigl(\theta+i\frac{\pi}{2},m_{\mathrm{eff}}(\vec{k}_{||})\bigr)=\frac{m_{\mathrm{eff}}(\vec{k}_{||})\cosh\theta-c_{j}}{m_{\mathrm{eff}}(\vec{k}_{||})\cosh\theta+c_{j}};\quad j=1,2.\]
 Introducing the variable $q$ by $m_{\mathcal{\mathrm{eff}}}(\vec{k}_{||})\cosh\theta=\sqrt{m^{2}+q^{2}}$
and performing the angular integrations one obtains \begin{eqnarray*}
E(L) & = & \frac{1}{(4\pi)^{D/2}\Gamma(D/2)}\int\limits _{0}^{\infty}dqq^{D-1}\\
 &  & \hspace{-1.6cm}\log\left(1-\frac{\sqrt{m^{2}+q^{2}}-c_{1}}{\sqrt{m^{2}+q^{2}}+c_{1}}\frac{\sqrt{m^{2}+q^{2}}-c_{2}}{\sqrt{m^{2}+q^{2}}+c_{2}}e^{-2L\sqrt{m^{2}+q^{2}}}\right).\end{eqnarray*}
 This formula is a new result of the present paper. To show its correctness
we consider two already known limiting cases. First consider the limit
when both $c_{j}\rightarrow0$ or $c_{j}\rightarrow\infty$ corresponding
to Neumann or Dirichlet boundary conditions for $\Phi$. In both cases
the coefficient of the exponent in the logarithm becomes one, and
we obtain the Ambjorn-Wolfram result for Dirichlet boundary conditions
\cite{ambjorn} as reported in Milton's book \cite{book}. In the
second limit we let the mass of the scalar field vanish $m\rightarrow0$;
then \begin{eqnarray*}
E(L) & =\\
 &  & \hspace{-2cm}\frac{1}{(4\pi)^{D/2}\Gamma(D/2)}\int\limits _{0}^{\infty}dqq^{D-1}\log\left(1-\frac{q-c_{1}}{q+c_{1}}\frac{q-c_{2}}{q+c_{2}}e^{-2Lq}\right).\end{eqnarray*}
 It is straightforward to show that this result coincides with that
obtained in \cite{robin} for the Casimir energy per unit area of
a \textsl{massless} scalar field with Robin boundary conditions.

The second application concerns the Casimir force between parallel
dielectric slabs separated by a vacuum slot of width $L$, a problem
first investigated by Lifshitz and collaborators almost fifty years
ago \cite{lifshitz}. In this geometry the dielectric constants in
the three regions are: \begin{eqnarray*}
\epsilon(x)=\epsilon_{1};\quad x<0; &  & \hspace{-1cm}\epsilon(x)=\epsilon_{2};\quad L<x;\\
 & \hspace{-.5cm}\epsilon(x)=1;\quad0<x<L.\end{eqnarray*}
 In this case $R_{1,2}^{\pm}(i\frac{\pi}{2}+\theta)$ are nothing
else but the (appropriate analytic continuations of the) ordinary
reflection amplitudes of electromagnetic waves incident from vacuum
at the plane interface between the vacuum and the dielectric materials,
given in many textbooks (e.g. \cite{Jackson}). Indeed denoting again
by $(k_{\perp},\vec{k_{||}})$, $\omega$ the wave vector and frequency
of the electromagnetic radiation ($\vec{k}_{||}$ being the component
parallel, while $k_{\perp}$ the one perpendicular to the plane interface)
the reflection amplitudes are \[
R_{\mathrm{perp}}^{(i)}(\omega,\vec{k}_{||})=\frac{\sqrt{\omega^{2}-\vec{k}_{||}^{2}}-\sqrt{\epsilon_{i}\omega^{2}-\vec{k}_{||}^{2}}}{\sqrt{\omega^{2}-\vec{k}_{||}^{2}}+\sqrt{\epsilon_{i}\omega^{2}-\vec{k}_{||}^{2}}},\quad\epsilon_{i}=\epsilon_{i}(\omega),\]
 when the electric field $\vec{E}$ is perpendicular to the plane
of incidence, and \[
R_{\mathrm{par}}^{(i)}(\omega,\vec{k}_{||})=\frac{\epsilon_{i}\sqrt{\omega^{2}-\vec{k}_{||}^{2}}-\sqrt{\epsilon_{i}\omega^{2}-\vec{k}_{||}^{2}}}{\epsilon_{i}\sqrt{\omega^{2}-\vec{k}_{||}^{2}}+\sqrt{\epsilon_{i}\omega^{2}-\vec{k}_{||}^{2}}},\quad\epsilon_{i}=\epsilon_{i}(\omega),\]
 when $\vec{E}$ is parallel to the plane of incidence and $i=1,2$.
(To obtain these expressions we assumed that the permeabilities of
the slabs are unity, $\mu_{i}=1$). Continuing $\theta\rightarrow\theta+i\frac{\pi}{2}$
corresponds to continuing $\omega$ to purely imaginary values $\omega\rightarrow m_{\mathrm{eff}}(\vec{k}_{||})i\sinh\theta=i\zeta,$
and the reflection amplitudes become ($q^{2}=\vec{k}_{||}^{2}$):
\begin{eqnarray*}
R_{\mathrm{perp}}^{(i)}(i\zeta,q) & = & \frac{\sqrt{\zeta^{2}+q^{2}}-\sqrt{\epsilon_{i}\zeta^{2}+q^{2}}}{\sqrt{\zeta^{2}+q^{2}}+\sqrt{\epsilon_{i}\zeta^{2}+q^{2}}},\quad\\
R_{\mathrm{par}}^{(i)}(i\zeta,q) & = & \frac{\epsilon_{i}\sqrt{\zeta^{2}+q^{2}}-\sqrt{\epsilon_{i}\zeta^{2}+q^{2}}}{\epsilon_{i}\sqrt{\zeta^{2}+q^{2}}+\sqrt{\epsilon_{i}\zeta^{2}+q^{2}}}.\end{eqnarray*}
 Since the electromagnetic field can have both polarizations, one
obtains the following form for the Casimir energy per unit area \begin{eqnarray*}
E(L) & = & \frac{1}{8\pi^{2}}\int\limits _{0}^{\infty}dq^{2}\int\limits _{0}^{\infty}d\zeta\Big[\\
 &  & \log\left(1-R_{\mathrm{perp}}^{(1)}(i\zeta,q)R_{\mathrm{perp}}^{(2)}(i\zeta,q)\mathrm{e}^{-2L\sqrt{q^{2}+\zeta^{2}}}\right)+\\
 &  & \log\left(1-R_{\mathrm{par}}^{(1)}(i\zeta,q)R_{\mathrm{par}}^{(2)}(i\zeta,q)\mathrm{e}^{-2L\sqrt{q^{2}+\zeta^{2}}}\right)\Big].\end{eqnarray*}
 The Casimir force ${\mathcal{F}}(L)=-\partial E(L)/\partial L$ computed
from this expression agrees with that of Lifshitz et al. as reported
in the reviews \cite{reviews,book}.

Next we consider the Casimir energy of a massless fermion field in
$1+3$ dimensions subject to the {}``bag boundary condition'' in
our planar geometry: \[
(1-i\gamma^{3})\psi|_{x=0}=0\quad,\qquad(1+i\gamma^{3})\psi|_{x=L}=0\quad.\]
 Using the chiral representation of Dirac matrices one readily shows
that these boundary conditions break chirality but commute with $\Sigma_{3}=\left(\begin{array}{cc}
\sigma_{3} & 0\\
0 & \sigma_{3}\end{array}\right)$. Furthermore they imply that for both the positive ($u$) and the
negative ($v$) eigenvectors of $\Sigma_{3}$ the reflection amplitudes
are constants, satisfying $R_{1(s)}^{-}(i\pi/2+\theta)R_{2(s)}^{+}(i\pi/2+\theta)=-1$
for $s=u,v$. Thus (taking into account crossing-unitarity (\ref{eq:crossing-unitarity}))\[
E(L)=-2\frac{1}{4\pi^{2}}\int\limits _{0}^{\infty}dqq^{2}\log(1+e^{-2Lq}),\]
 which indeed coincides with the known result \cite{book,fermion}.

\section{Concluding remarks}

The main result of the paper is formula (\ref{eq:leading_order})
which expresses the Casimir energy in large volume in a general form
valid for any QFT - even for theories interacting in the bulk - in
terms of the reflection amplitudes (infrared data). It passed the
test of interacting but integrable two dimensional quantum field theories
in \cite{bluscher}. The boundary state approach provides a systematic
infrared (large volume) expansion, which in the case of noninteracting
bulk and elastic boundary scattering can be summed up leading to (\ref{eq:alt3}).
In order to check the result for noninteracting bulk we present in
the appendix an alternative derivation using the naive mode summation
method. 

In the usual calculations of the Casimir effect corrections for temperature
and roughness effects must be made before comparing to the experiment.
Both corrections can be introduced into (\ref{eq:leading_order},\ref{eq:alt3})
in a straightforward manner, following the usual procedure in the
literature (cf. \cite{book}) see also \cite{rough}. In the case
of a massive theory with mass gap $m$ the terms coming from higher
particle corrections to (\ref{eq:B_expansion}) and (\ref{eq:D_expansion})
are suppressed by $\exp(-mL)$. For theories with a zero mass gap
the situation is different and indeed the asymptotic state formalism
is not well-founded if interaction with massless particles (such as
photons) is taken into account, due to infrared divergencies. In the
case of electromagnetic field the corrections to (\ref{eq:alt3})
are known to be suppressed by $\alpha\lambda/L$ where $\lambda$
is the Compton wave length of the electron and $\alpha$ is the fine
structure constant (Chapter 13 of \cite{book}), due to the fact that
the self-interaction of the electromagnetic field arises only through
radiative corrections and the leading one is an electron loop. 

The striking fact about the general formulae (\ref{eq:leading_order},\ref{eq:alt3})
is that they are manifestly finite, and only depend on physical quantities,
such as the dispersion relation of asymptotic particles and their
reflection amplitudes on the boundaries. This is even more emphasized
in the boundary state formalism, where the bulk and boundary divergences
appearing in usual derivations of the Casimir effect are manifestly
absent, and the whole derivation is performed in terms of finite (renormalized)
physical objects. The appearance of divergences in the naive method
of summing zero-point energies is a general feature when one calculates
the interaction energy of point-like sources by integrating the energy
stored in the field generated by them. The divergences are then eliminated
by noticing that they contribute to the self-energy of the sources
themselves and are present even at infinite separation. These divergences
are absorbed by renormalization, but they are entirely unphysical
and eliminated when expressing the measurable energy differences in
terms of physically meaningful, finite quantities. Further bonus of
the general formulae (\ref{eq:leading_order},\ref{eq:alt3}) is that
the reflection factors appearing in them can be calculated easily
in the semi-infinite geometry, e.g. using perturbation theory along
the lines of \cite{bLSZ}.

\section*{Acknowledgments}

This research was partially supported by the EC network {}``EUCLID'',
contract number HPRN-CT-2002-00325, and Hungarian research funds OTKA
D42209, T037674, T043582 and TS044839. GT was also supported by a
Sz{\'e}chenyi Istv{\'a}n Fellowship.

\appendix

\section{Casimir energy by mode summation}

Here we want to support our derivation coming from the boundary state
formalism by obtaining the analogous result using a new version of
the mode summation method for fluctuating fields which are free in
the bulk. Ideas somewhat similar to these have been used in a similar
framework earlier in \cite{JR} and their findings also support our
general expression. The novel viewpoints of our approach are twofold:
first we concentrate only on the finite size piece of the ground state
energy that should be independent of regularization and renormalization
thus we do not need to discuss the regularization/renormalization
of the (infinite space) bulk/boundary ground state energies, since
eventually these are subtracted. The second feature of our approach
is that we do not solve the quantization condition for the zero point
frequencies explicitly but use only the functional form of these equations
to obtain a general expression for the ground state energy containing
only the reflection amplitudes. 

For simplicity let us start with the $1+1$ dimensional free scalar
field $\Phi(t,x)$ of mass $m$ in a domain described earlier. We
assume that in all three domains the frequency of the modes of $\Phi$
is the same, but the wave vectors $k'$, $k''$ may be different from
$k$ (e.g. the mass of the scalar field is different in the three
domains), but they can be regarded as functions of $k$. All the wave
number parameters $k,\, k',\, k''$ are chosen positive. Then for
the plane wave modes one has the situation displayed in the following
figure 

\medskip{}
\hspace{-.5cm}\includegraphics[%
  width=9cm,
  keepaspectratio]{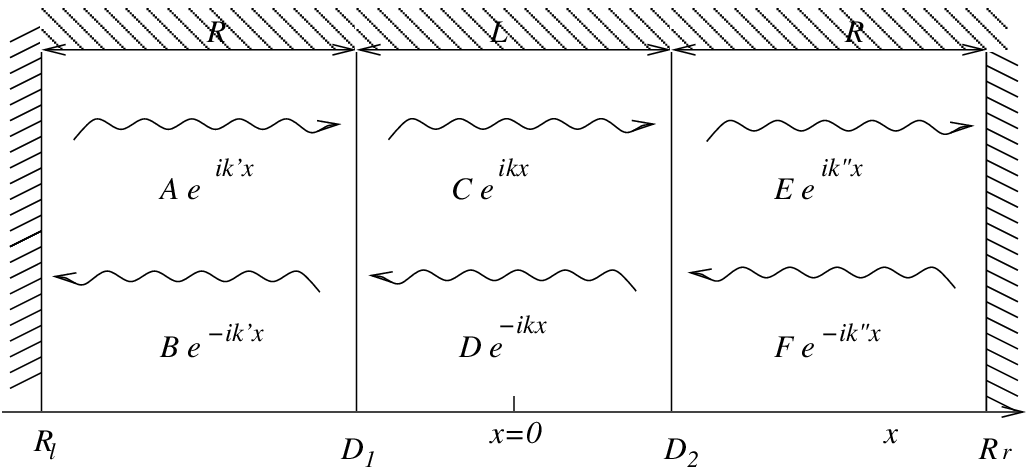}

where total reflection occurs at $x_{l}=-R-L/2$ and $x_{r}=R+L/2$
with reflection factors $R_{l}$ and $R_{r}$ 

\begin{eqnarray}
Ae^{-ik'(R+\frac{L}{2})} & = & Be^{ik'(R+\frac{L}{2})}R_{l}(k')\label{refl}\\
Fe^{-ik''(R+\frac{L}{2})} & = & Ee^{ik''(R+\frac{L}{2})}R_{r}(k''),\nonumber \end{eqnarray}
and the two {}``defects'' (i.e. boundaries allowing transmission)
located at $x_{1}=-L/2$ and $x_{2}=L/2$ describe one-particle scattering
with defect matrices \[
D_{i}(k)=\left(\begin{array}{cc}
R_{i}^{+}(k) & T_{i}^{-}(k)\\
T_{i}^{+}(k) & R_{i}^{-}(k)\end{array}\right),\qquad i=1,2\]
In the case of elastic defect scattering their is no particle production
and the defect matrices are unitary. They connect the incoming and
outgoing amplitudes as follows\begin{eqnarray}
\left(\begin{array}{c}
B\mathrm{e}^{+ik'L/2}\\
C\mathrm{e}^{-ikL/2}\end{array}\right) & = & D_{1}(k)\left(\begin{array}{c}
A\mathrm{e}^{-ik'L/2}\\
D\mathrm{e}^{+ikL/2}\end{array}\right)\label{eq:defect}\\
\left(\begin{array}{c}
D\mathrm{e}^{-ikL/2}\\
E\mathrm{e}^{+ik''L/2}\end{array}\right) & = & D_{2}(k)\left(\begin{array}{c}
C\mathrm{e}^{+ikL/2}\\
F\mathrm{e}^{-ik''L/2}\end{array}\right)\nonumber \end{eqnarray}
Unitarity and time-reversal gives the relations \begin{eqnarray}
D_{i}(k)^{\dagger}D_{i}(k)=1 & , & D_{i}(k)^{-1}=D_{i}(-k)\label{eq:unitarity}\\
R_{r,l}(k)^{*}R_{r,l}(k)=1 & , & R_{r,l}(k)^{-1}=R_{r,l}(-k)\quad.\nonumber \end{eqnarray}
Consistency of the homogenous linear system (\ref{refl},\ref{eq:defect})
is expressed by the quantization condition\begin{equation}
Q(k_{n})=0\label{eq:quantization}\end{equation}
where{\small \begin{eqnarray*}
Q(k)=\bigl(1-R_{l}(k)R_{1}^{+}(k)e^{i2k'R}\bigr)\bigl(1-R_{r}(k)R_{2}^{-}(k)e^{i2k''R}\bigr)-\\
e^{i2kL}R_{1}^{-}(k)R_{2}^{+}(k)\biggl(1-\frac{R_{l}(k)e^{i2k'R}}{R_{1}^{+}(-k)}\biggr)\biggl(1-\frac{R_{r}(k)e^{i2k''R}}{R_{2}^{-}(-k)}\biggr)\end{eqnarray*}
}and we exploited (\ref{eq:unitarity}) to write \[
\det{D_{i}}=R_{i}^{-}(k)/R_{i}^{+}(-k)=R_{i}^{+}(k)/R_{i}^{-}(-k)\]
Due to unitarity the quantization condition can only have real and
purely imaginary solutions. For each real solution $k_{n}$, $-k_{n}$
is also a solution, and $k_{0}=0$ always solves (\ref{eq:quantization}).
The purely imaginary solutions located at $k_{j}=i\kappa_{j}$ ($0<\kappa_{j}<m$)
are related to the poles of the reflection and transmissions factors
and correspond to defect and boundary bound states. Then the ground
state energy of the system is given by the sum of zero modes, which
can be written in an integral form\begin{eqnarray*}
 &  & E(L,R)=\sum_{\{ k_{n}>0\}}\frac{1}{2}\omega(k_{n})+\sum_{j}\frac{1}{2}\omega(i\kappa_{j})=\\
 &  & \frac{1}{2}\left\{ \frac{1}{2}\left(\int_{C_{+}}+\int_{C_{-}}-\int_{C_{0}}\right)+\sum_{j}\oint_{C_{j}}\right\} \frac{dk}{2\pi i}\frac{Q'(k)}{Q(k)}\omega(k)\end{eqnarray*}
 where $Q'(k)=\frac{dQ}{dk}$, $\omega(k)=\sqrt{k^{2}+m^{2}}$, $\hbar$
is set to one and the contours are shown on the figure

\begin{center}\hspace{-.4cm}\includegraphics[%
  width=8.5cm,
  keepaspectratio]{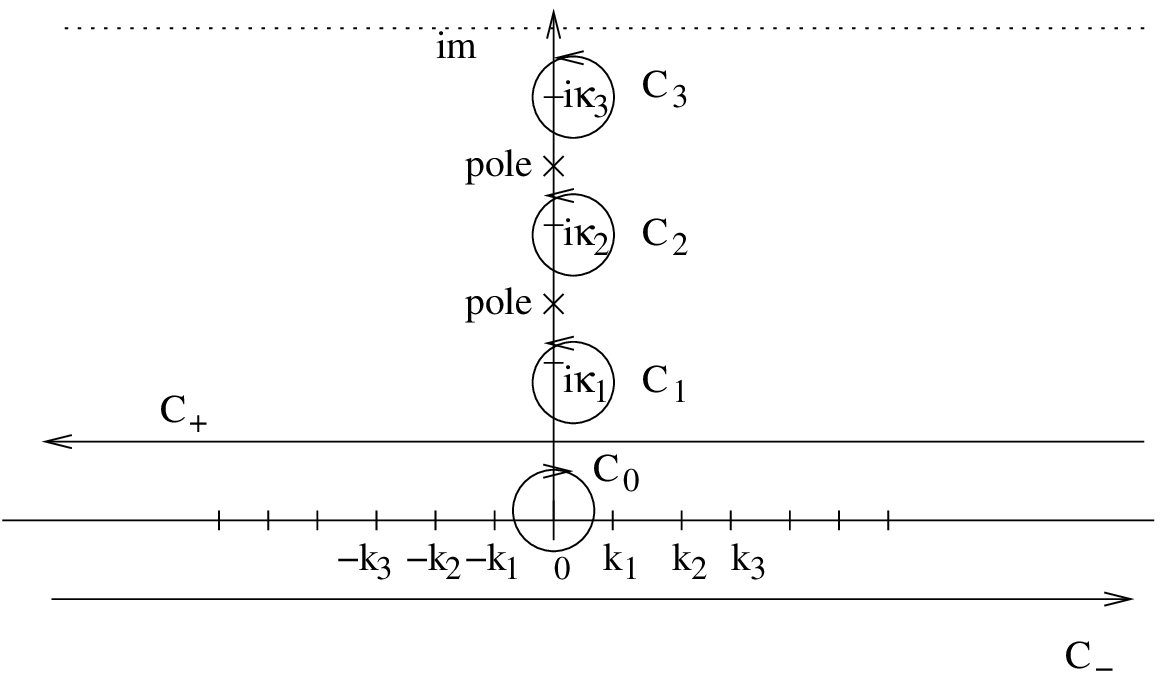}\end{center}

The poles on the figure denote the poles of the $Q$ function originating
from the poles of the reflection and transmission amplitudes. Contrary
to $\kappa_{j}$, which do depend on the volume, the location of these
poles is independent of $L,\, R$. 

We concentrate on the $L$ dependence so the $L$ independent $C_{0}$
integral can be dropped. Next we analyze the contribution of $C_{-}$.
Using (\ref{eq:unitarity}) we can relate\begin{eqnarray*}
\frac{Q'(-k)}{Q(-k)}-\frac{Q'(k)}{Q(k)} & = & -2iL-2iR\left(\frac{dk'}{dk}+\frac{dk''}{dk}\right)\\
 &  & \hspace{-2cm}-\frac{d}{dk}\log R_{r}-\frac{d}{dk}\log(\det{D}_{1}{D}_{2})-\frac{d}{dk}\log R_{l}.\end{eqnarray*}
 The first two terms on the right hand side (proportional to the volumes
$L$, $R$) correspond to divergent bulk energy contributions. The
other terms here correspond to infinite boundary and defect energies
which are independent of the volumes ($L$, $R$). The precise definition
of these terms needs some regularization and renormalization, but
they give no contribution to the Casimir force since they are present
in the infinite system as well. Therefore they can be dropped and
then the $C_{+}$ and $C_{-}$ integral terms give identical results.

Thus the relevant contribution to the Casimir energy can be written
in terms of the variable $m\sinh\theta=k$ as\[
E(L,R)=-m\int_{-\infty+i\eta}^{\infty+i\eta}\frac{d\theta}{4\pi i}\,\cosh\theta\:\frac{Q'(\theta)}{Q(\theta)}+\sum_{j}\frac{1}{2}\omega(i\kappa_{j}).\]
(where $\eta$ is a small positive parameter). Next we shift the contour
to $\eta=\frac{\pi}{2}$ which eliminates the discrete sum via the
contributions from the purely imaginary zeros of $Q(\theta)$. There
can be additional contributions from poles of $Q\left(\theta\right)$
on the imaginary axis, but their positions are independent of the
volume and therefore they only contribute volume independent terms
which can be dropped. Integrating by parts and introducing the new
variable $\theta\rightarrow\frac{i\pi}{2}+\theta$ we obtain \[
E(L,R)=m\int_{-\infty}^{\infty}\frac{d\theta}{4\pi}\,\cosh\theta\:\log Q(\frac{i\pi}{2}+\theta).\]
In the limit $R\rightarrow\infty$\begin{eqnarray*}
E(L) & = & m\int_{-\infty}^{\infty}\frac{d\theta}{4\pi}\,\cosh\theta\:\\
 &  & \hspace{-1cm}\log\left(1-R_{1}^{-}\left(\frac{i\pi}{2}+\theta\right)R_{2}^{+}\left(\frac{i\pi}{2}+\theta\right)\mathrm{e}^{-2m\cosh\theta L}\right).\end{eqnarray*}
It is straightforward to generalize these expressions to the $D+1$
dimensional case when the boundary conditions are specified on two
parallel $D-1$ dimensional hyperplanes. Introducing the component
parallel ($\vec{k}_{||}$) respectively perpendicular ($k_{\perp}$)
to the hyperplanes one can write \begin{eqnarray}
\vec{k} & = & (k_{\perp},\vec{k}_{||});\quad\omega^{2}-k_{\perp}^{2}=m^{2}+\vec{k}_{||}^{2}\equiv m_{\mathrm{eff}}^{2}(\vec{k}_{||});\nonumber \\
 &  & k_{\perp}=m_{\mathrm{eff}}(\vec{k}_{||})\sinh\theta.\label{eq:rapidity}\end{eqnarray}
We assume that the system is translationally invariant in the direction
of the hyperplanes, and so the reflection and transmission processes
preserve $\vec{k}_{||}$. Therefore we can apply the previous considerations
to the decoupled one-dimensional systems labeled by the parameter
$\vec{k}_{||}$ and sum up their contributions. Note that both $\omega$
and the reflection/transmission amplitudes depend on $\vec{k}_{||}$,
but we shall not write this out explicitely. The Casimir energy per
unit transverse area can be written as\begin{eqnarray}
E(L,R) & =\label{alt}\\
 &  & \hspace{-2cm}\int\frac{d^{D-1}\vec{k}_{||}}{(2\pi)^{D-1}}\, m_{\mathrm{eff}}(\vec{k}_{||})\int_{-\infty}^{\infty}\frac{d\theta}{4\pi}\,\cosh\theta\:\log Q\left(\frac{i\pi}{2}+\theta\right).\nonumber \end{eqnarray}
From this, the Casimir energy between the two defects can be obtained
by taking the limit $R\rightarrow\infty$\begin{eqnarray}
E(L) & = & \int\frac{d^{D-1}\vec{k}_{||}}{(2\pi)^{D-1}}m_{\mathrm{eff}}(\vec{k}_{||})\int_{-\infty}^{\infty}\frac{d\theta}{4\pi}\,\cosh\theta\:\label{alt2}\\
 &  & \hspace{-1.6cm}\log\left(1-R_{1}^{-}\left(\frac{i\pi}{2}+\theta\right)R_{2}^{+}\left(\frac{i\pi}{2}+\theta\right)\mathrm{e}^{-2m_{\mathrm{eff}}(\vec{k}_{||})\cosh\theta L}\right).\nonumber \end{eqnarray}
Note that the latter result is independent of the auxiliary boundary
conditions $R_{r}$ and $R_{l}$ which can be expected on physical
grounds. Taking into account the crossing-unitarity property satisfied
by the reflection amplitudes \cite{GZ,defect} which for free bosons/fermions
takes the form\begin{equation}
R\left(\frac{i\pi}{2}+\theta\right)=\pm R\left(\frac{i\pi}{2}-\theta\right)\label{eq:crossing-unitarity}\end{equation}
one can see that this result is in a complete agreement with the formula
obtained form the boundary state formalism previously (\ref{eq:alt3}).
(The results in (\ref{alt}, \ref{alt2}) should be multiplied by
$(-1)$ in case of computing the Casimir energy of fermionic fields,
since their vacuum energy is $-\frac{1}{2}\sum\omega(k)$).

If there are more than one fields in the problem, or the fields have
more than one component one has to sum up the contribution of all
the fields (components) to obtain the Casimir energy.

\end{document}